\documentclass[%
nofootinbib,
 amsmath,amssymb,
 aps,
prf,
onecolumn,
]
{revtex4-2}

\usepackage{graphicx}
\usepackage{dcolumn}
\usepackage{bm}
\usepackage{xcolor}

\begin{document}

\title{Buoyancy driven dissolution of inclined blocks: Erosion rate and pattern formation}

\author{Caroline Cohen}
\author{Michael Berhanu}%
\author{Julien Derr}%
\author{Sylvain Courrech du Pont}%
 \email{sylvain.courrech@univ-paris-diderot.fr}
\affiliation{%
Laboratoire Mati\`eres et Syst\`emes Complexes, Universit\'e de Paris, CNRS, \\B\^{a}timent Condorcet, 10 rue Alice Domon et L\'eonie Duquet, 75205 Paris cedex 13, France\\
}%





\begin{abstract}
The dissolution of a body into quiescent water leads to density stratifications at the interfaces that drive buoyant flows. Where the stratification is unstable, the flow destabilizes into convective solute plumes. By analogy with the Rayleigh-B\'enard instability where concentration replaces temperature, this phenomenon is known as the solutal Rayleigh-B\' enard instability. Here we report experiments of the dissolution of inclined rectangular blocks made of salt, caramel or plaster in aqueous solutions of various concentrations. The solute flows along the block while forming plumes before they detach and sink. This flow along the block organizes the emission of plumes within longitudinal parallel stripes with a well-defined millimeter scale wavelength. The instability of the flow reflects on the concentration field in the boundary layer, which engraves longitudinal grooves onto the block. These grooves interact with the flow and turn into a paving of 3 dimensional cup like patterns that grow in size and propagate upstream. These bedforms are reminiscent of the scallop bedforms observed on the walls of cave or icebergs. Whereas the block interface is highly dynamical and evolves through time, it remains flat on the global scale and recedes at a stationary rate. We derive scaling laws for the receding velocity and the pattern genesis at the inclined interface that are based on a concentration boundary layer of constant thickness, which is controlled by the flow instability but where neither the patterns nor the flow along the block do not play any role. We apply these results to the formation of sublimation patterns.

\end{abstract}

\maketitle


\section{\label{}Introduction}

A water flow, and its interaction with a topography, modifies the rate of phase transition, or of dissolution or precipitation of a body. In nature, the coupling between the geometry and the mass transfer drives the formation of recognizable patterns at different scales. These can be erosion patterns like river meanders, cyclic steps, rillenkarren, lapiaz, ice ripples, dissolution flutes and scallops, regmaglypts on meteorites or deposition shapes like travertines, stalactites, icicles or brinicles. These patterns are not geological curiosities only but are markers of the hydrodynamic processes, which often control the global erosion process. The identification of erosion and deposition patterns is therefore a key to infer flow and thermodynamic conditions. It is especially valuable because these structures are quite resilient and give information on the long term not only for Earth but also for other planetary bodies that are difficult to instrument.

Here we study the dissolution of rectangular blocks immersed in still aqueous solutions in the case where the solute stratification it engenders is buoyancy unstable \cite{Cohen_2016}. The concentration boundary layer destabilizes to form plumes that drive a convective flow of solute. This heterogeneous flow influences the concentration field at the interface, which patterns the dissolving body. 

Several studies address the growth or dissolution of solids into aqueous solutions and the subsequent turbulent solutal convective flow \cite{Schurr_1905, Garner_1961, Thomas_1968, Grijseels_1981, Hurle_1982, Tait_1989, Kerr_1994a, Sullivan_1996, Wykes_2018}. Most of them do the parallel with the Rayleigh-B\' enard instability where the buoyancy is not due to the thermal expansion of the liquid but to the gradients of solute concentration. These studies are either motivated by industrial applications from a chemical engineering perspective \cite{Grijseels_1981, Hurle_1982} or by geophysical questions such as the sequestration of $\rm{CO_2}$ in porous rocks \cite{Neufeld_2010, Huppert_2014, Loodts_2014, Slim_2014} or the formation of patterns like brinicles or corrugations on the ice cap of brackish lakes or in magma chambers \cite{Tait_1989, Tait_1992a, Kerr_1994a, Meakin_2009, Solari_2013}. 

Our experimental study stands out from these previous studies by addressing the dissolution of inclined blocks. The inclination of the block gives a direction to the dissolution flow. As a result, the instability of the flow and the primary patterns are longitudinal with a well-defined wavelength. The latter interplay between the dissolution flow and the shaped topography gives rise to dynamical three dimensional structures that propagate upstream the dissolution flow.
In the next section (sec. \ref{SecExp}) we detail our experiments with caramel, salt or plaster blocks, which are fast dissolving materials. The caramel being transparent, we can follow the pattern formation in real time. In section \ref{SecScal} we derive scaling laws for the dissolution rate and for the characteristic time and wavelength of pattern formation in the case of inclined bodies. We compare the scaling laws to the experimental data in section \ref{SecRes}. We show in particular that the solute flow along the block and the dynamics of patterns at the block interface have little if any influence on the global dissolution rate.
Finally we put our work in perspective in a geomorphological context. In section \ref{SecPenitentes} we make the connection between the dissolution and sublimation patterns such as penitentes. In conclusion we link our work to the dissolution cavities observed in nature and compare the free dissolution patterns we observe in our experiments with their counterparts shaped by an externally imposed stream.

\section{Experiments \label{SecExp}}
\begin{figure*}[t]
\begin{center}
\includegraphics[width =\linewidth]{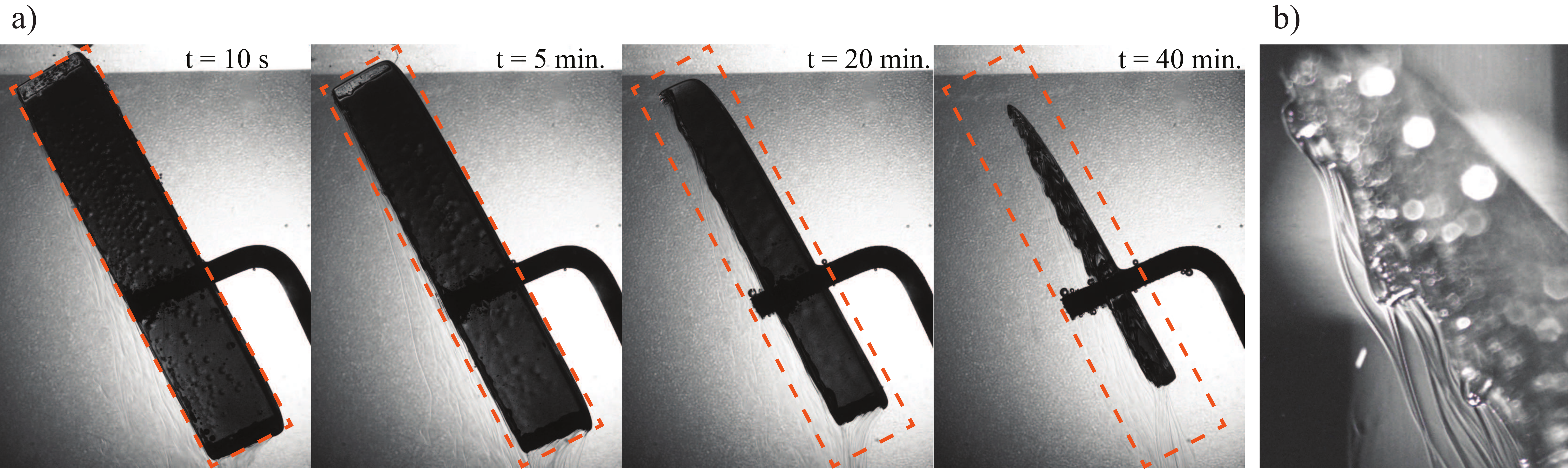}
\caption{a) Side view of a caramel block dissolving into water. The initial block is $6 \ {\rm{cm}}$-long and makes an angle of $62^{\circ}$ with the horizontal. Except in the vicinity of the top corner, the overall inclination of the receding bottom interface does not change over time. On the contrary, the top interface dissolves faster at the top (upstream) than at the bottom (downstream). b) Detail showing solute threads detaching from tips at the bottom interface.}
\label{FigCarCote}
\end{center}
\end{figure*}
\begin{figure*}[t]
\begin{center}
\includegraphics[width =\linewidth]{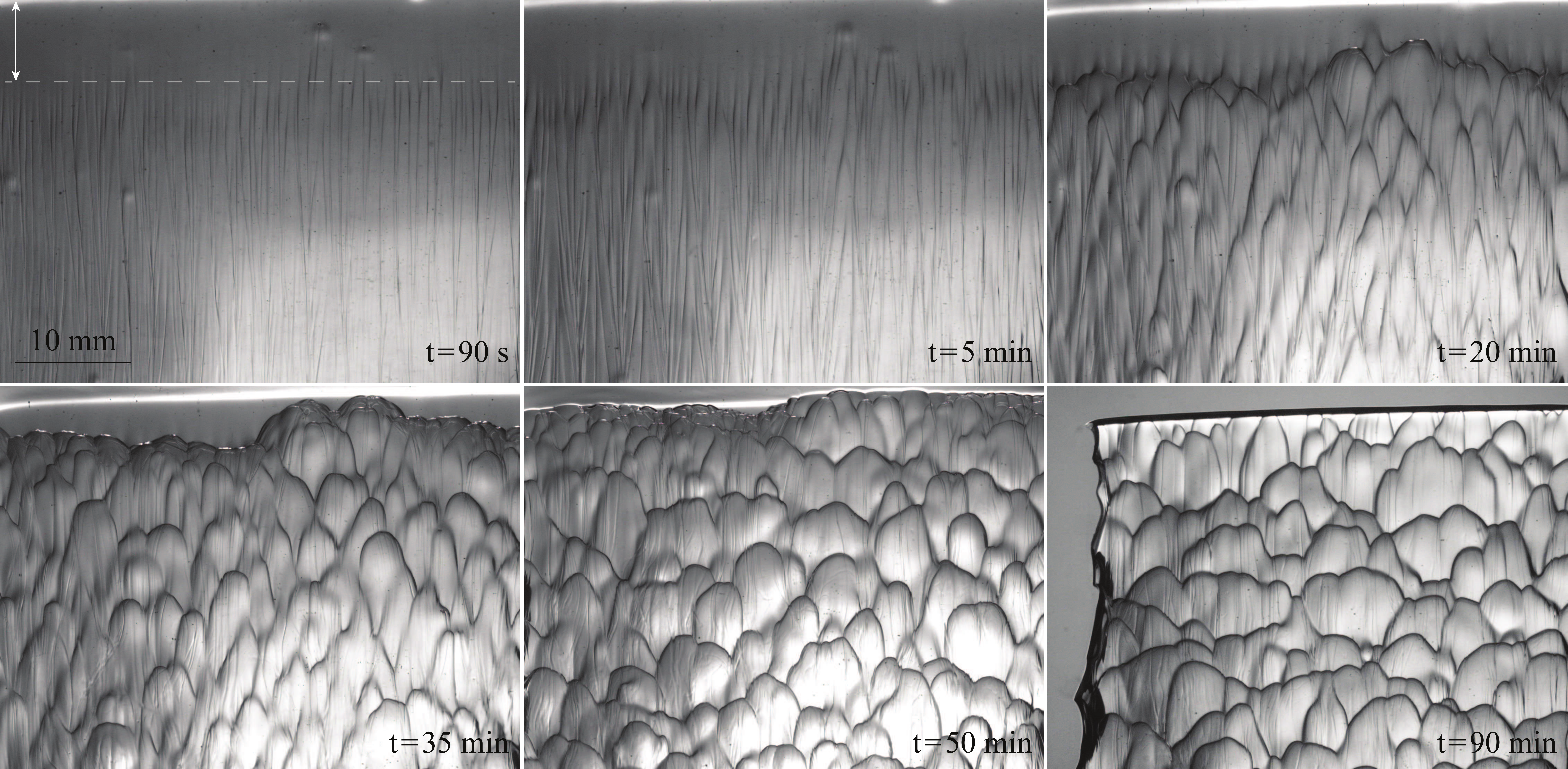}
\caption{Pattern formation. Bottom view of a caramel block dissolving into water. The block makes an angle of $60^{\circ}$ with the horizontal. The top of the first picture corresponds to the block end. The block is lit from the back, crests and peaks appear dark. Parallel stripes rapidly form beyond an entry length from the block end. Then stripes cross and evolve into scallops that propagate upstream. The white dashed line and arrows show the entry length on the first picture.}
\label{FigSeqCar}
\end{center}
\end{figure*}
\begin{figure}[t]
  \includegraphics[width =0.48\linewidth]{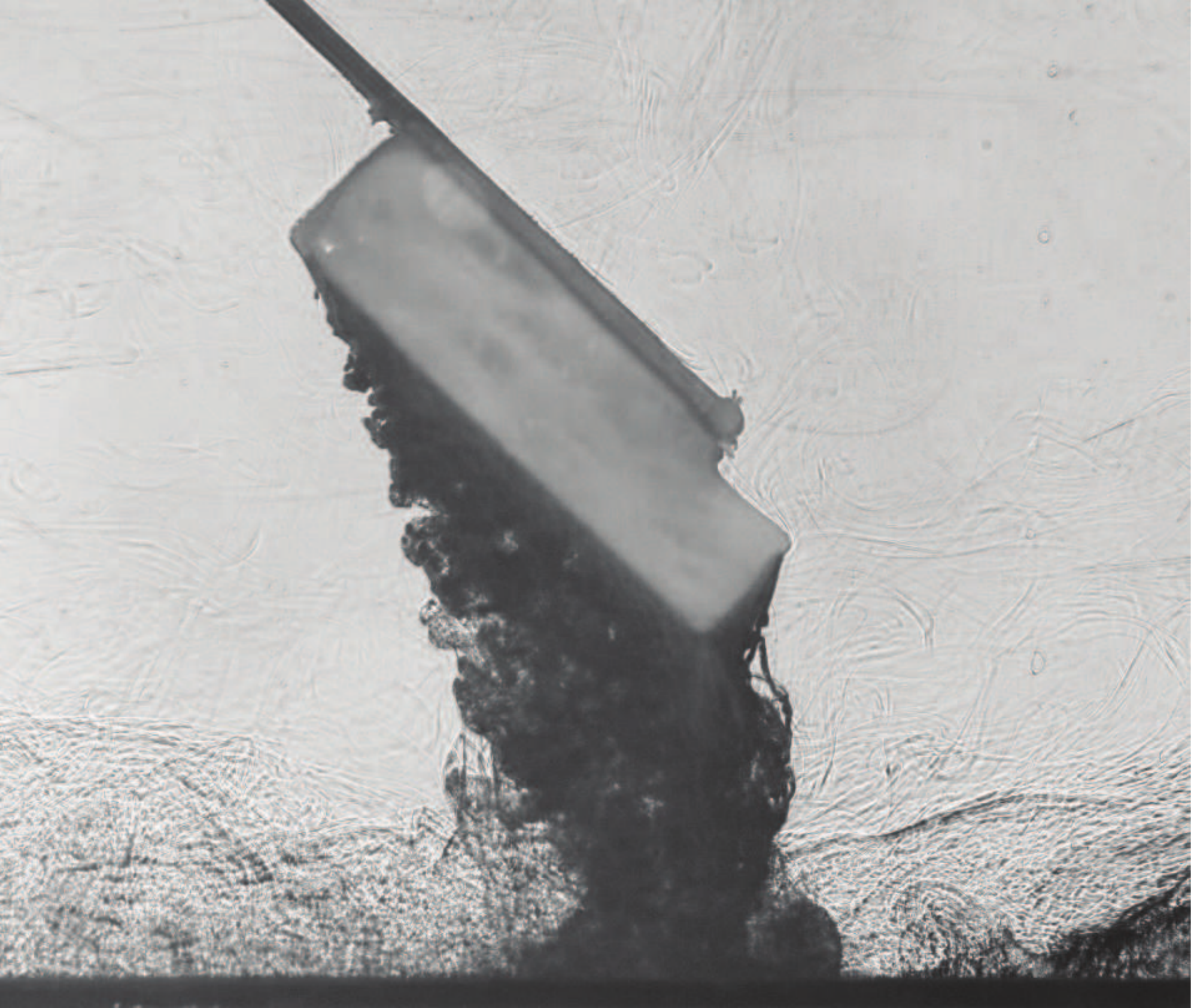}
\hfill
 \includegraphics[width =0.48\linewidth]{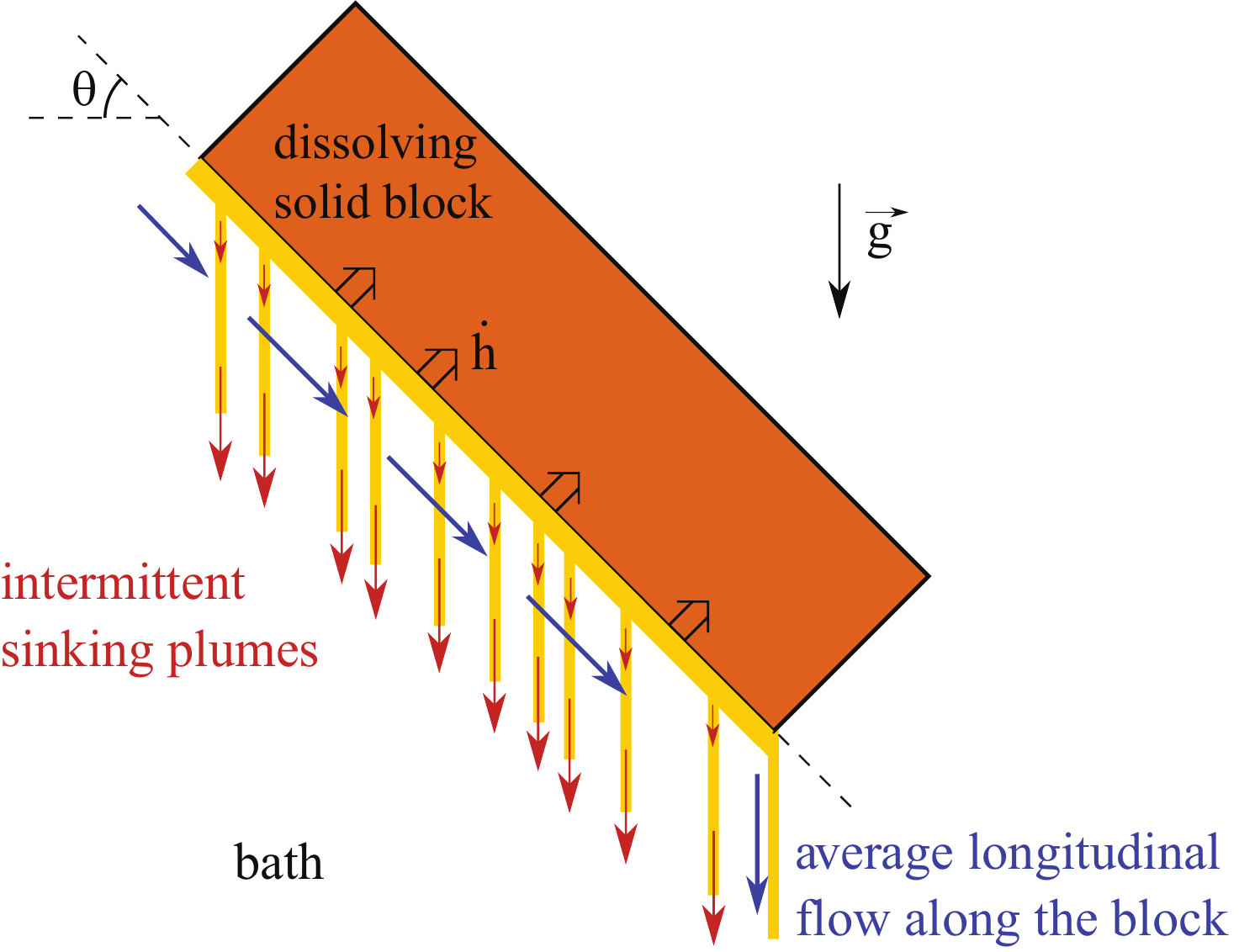}
 \caption{Left: Schlieren picture of a salt block dissolving into water. The block is $9 \, {\rm{cm}}$ long and makes an angle of $45^{\circ}$ with the horizontal. One sees the very different solute flows on the interfaces facing upwards and downwards. On the bottom interface, the solute-laden fluid also flows along the block before detaching because the block is inclined. This flow has some inertia as observed at the bottom corner of the block. One also sees that the  solute-laden fluid poorly mixes with the bath solution but sink and accumulate at the bottom of the tank. Right: Schematic view of the dissolution-driven flow.  
\label{FigSchlieren}} 
     \end{figure}
 We investigate the dissolution of rectangular blocks of either caramel, salt (NaCl) or plaster (gypsum) immersed in quiescent aqueous solutions.
 We systematically vary the inclination of the block and the solute concentration of the solution.

Salt blocks come from Himalayan mines. They are typically $10$ cm long, $5$ cm wide and $2.5$ cm thick and have a density of $\rho_{\rm{s}}=2348 \pm 5 \, \rm{kg. m^{-3}}$. We stick to blocks aluminum lugs with silicon sealant to hold them into the saline solution. We mix tap water and table salt to prepare solutions of various concentrations. The measured density of the saturated brine is  $\rho_{\rm{sat}}=1197.3 \pm 1 \, \rm{kg. m^{-3}}$ at $22^{\circ}\rm{C}$.  Caramel is made with saccharose, tap water and a few drops of lemon juice (acid) to favor the caramelization reaction. We cast hot caramel into silicon molds of $ 10\times 10 \times 3 $ cm or $ 5.5 \times 2.5 \times 2 $ cm to obtain rectangular blocks. An allen wrench is embedded in the block to hold it in the solution. The color of caramel is controlled during cooking to ensure that blocks are of similar properties and constant density ($\rho_{\rm{s}}=1540 \pm 10 \, \rm{kg.m^{-3}}$). We either mix tap water and caramel or tap water and saccharose to prepare the syrups. The measured density of the saturated caramel solution is $\rho_{\rm{sat}}=1450 \pm 10 \, \rm{kg.m^{-3}}$. We made the plaster block by pouring a $ 5.5 \times 2.5 \times 2 $ cm silicon mold with a mix of fine powder of plaster of Paris and water. The density of the block is $\rho_{\rm{s}} = 1859 \pm 5 \, \rm{kg. m^{-3}}$ and we measured a density difference of $1.8 \pm 0.6 \ \rm{kg. m^{-3}}$ between the saturated solution and water.

An experiment consists in immersing a block at room temperature (between 19 and $24 ^{\circ}\rm{C}$) at the center of a glass tank, which is $40$ cm long, $20$ cm wide and is filled with the solution to a depth of $20$ cm. We capture the receding of the bottom and the top interfaces with a camera (see fig. \ref{FigCarCote}). Caramel blocks are transparent so that we can also record the formation and the evolution of patterns on the bottom interface. To do so, we light the block from behind and either take pictures of the bottom interface thanks to an inclined mirror put into the tank or take frontal pictures with the camera positioned underneath and the tank tilted of the same angle as the immersed block (fig. \ref{FigSeqCar}). We cannot observe the patterns on the salt and plaster blocks when immersed. They are opaque and interfaces appear blurry because of the gradient of the solute concentration and consequently of the refractive index of the solution in the vicinity of interfaces. We take salt and plaster blocks out of solutions to observe the patterns.\\

\begin{figure}
\begin{center}
  \begin{minipage}{0.48\linewidth}
   \includegraphics[width=0.95\textwidth]{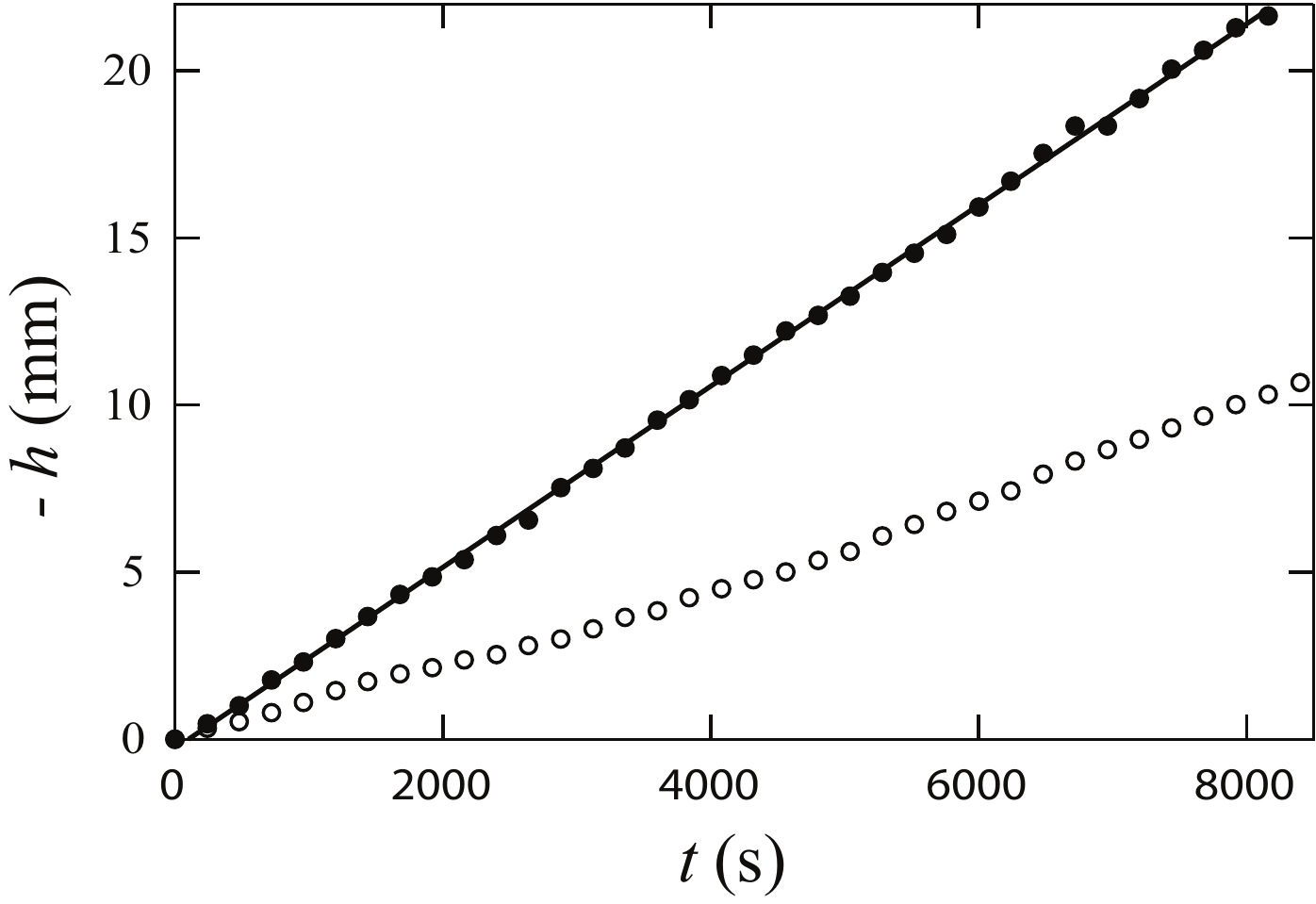}
   \end{minipage}\hfill
   \begin{minipage}{0.48\linewidth}
   \caption{Position $h$ taken in the middle of the bottom interface (plain points) and of the top interface (empty points) as a function of time $t$ for a caramel block inclined of $62^{\circ}$ dissolving into water. The black line shows a linear fit.  We measure positions along lines normal to the initial interfaces. We average the receding velocities from the measurements along five different lines along the block. The mean receding velocity of the bottom interface is $\dot{h}= - 2.78 \times 10^{-3} \pm 3 \times 10^{-5} \, {\rm mm/s}$. \label{FigHTCaramel}} 
   \end{minipage}
   \end{center}
\end{figure}
\begin{figure*}
\begin{center}
\includegraphics[width =\linewidth] {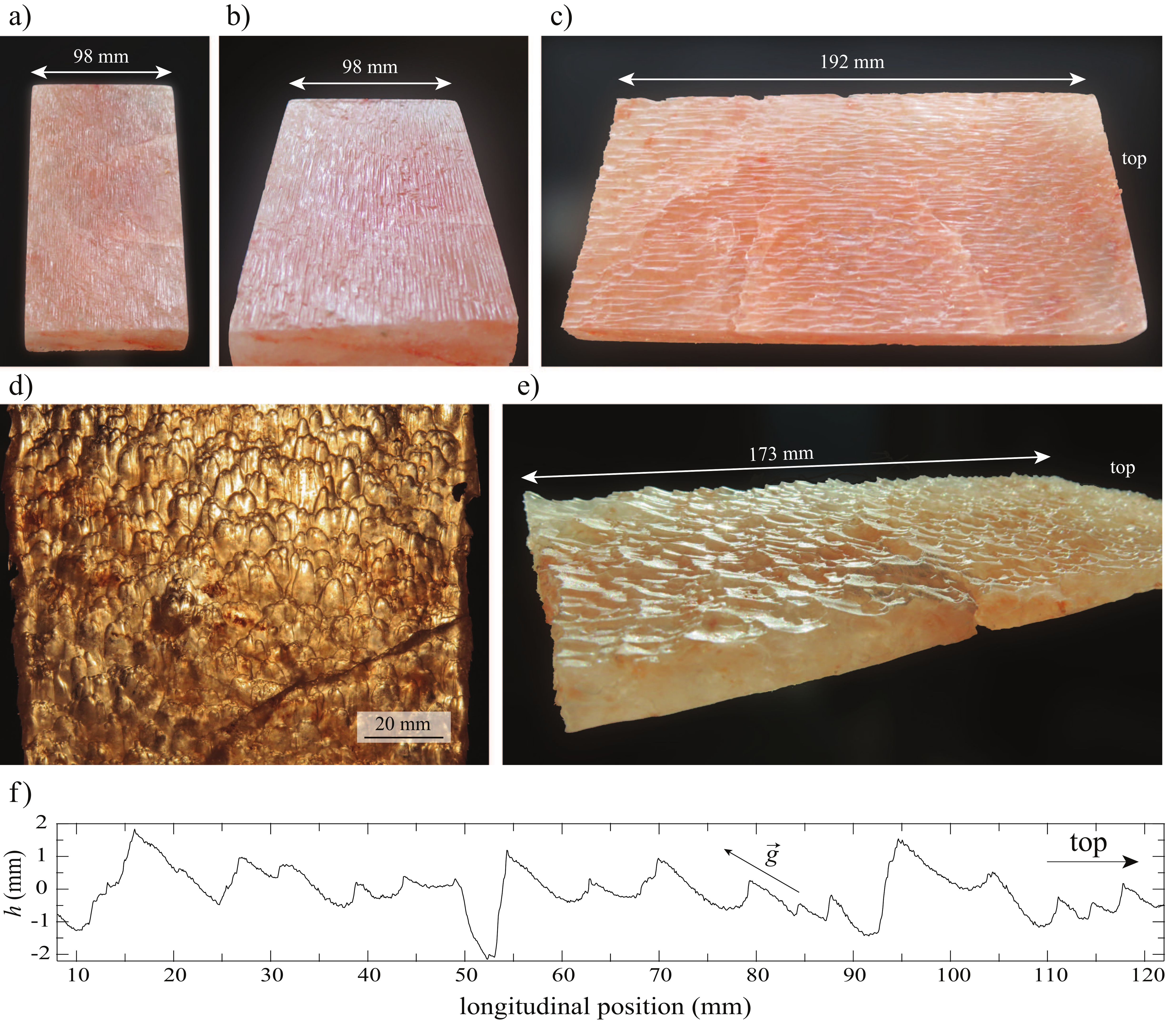}
\caption{Pattern formation. Bottom view of a salt block after 5 min. (a), 10 min. (b), 20 min. (c) and 1 hour (d-e) of immersion into pure water with an inclination of $60^{\circ}$ with the horizontal. Image (d) is lit from the back. Image (e) shows the asymmetry and pointy tips of dissolution patterns. Graphic (f) shows a line profile along this block. The slope on the stoss side seems parallel to the gravity in the vicinity of the tip. After an hour of immersion, the standard deviation of the topography is $0.62 \ {\rm{mm}}$.}
\label{FigSeqSel}
\end{center}
\end{figure*}
\begin{figure*}[t]
\begin{center}
\includegraphics[width =\linewidth]{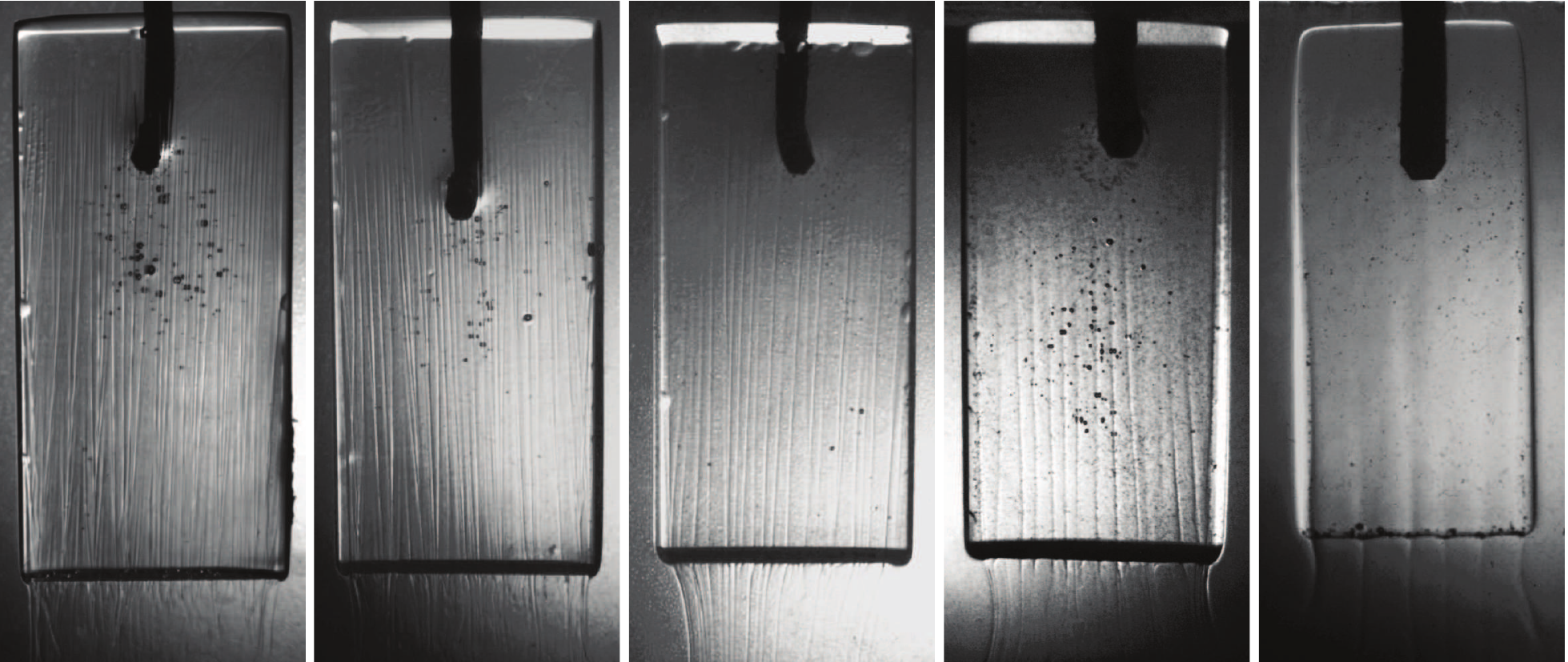}
\caption{Bottom view of caramel blocks ($55 \times 25 \, {\rm{mm}}$) dissolving into aqueous solutions of different caramel concentration. From left to right, the densities of solutions are $\rho_{\rm{b}} = {\rm{1006, \, 1108, \, 1229, \, 1303\, and \, 1403 \, kg.m^{-3}}}$. The inclination of blocks ranges between $70^{\circ}$ and $75^{\circ}$.}
\label{FigPar}
\end{center}
\end{figure*}

Figure \ref{FigCarCote} and \ref{FigSeqCar} show side views and bottom views of the evolution of an inclined caramel block immersed in water. The block dissolves, interfaces recede and we observe different flows and block dynamics depending on the orientation of the block walls. Threads of solute-laden fluid escape from the block interfaces pointing downwards and sink while the solute flows all along the block walls pointing upwards (fig. \ref{FigSchlieren}). The inclination of the block enforces a downward direction to the flow on either interface. On the bottom interface, the solute flows downwards before detaching (see fig. \ref{FigSchlieren}). Using Particle Image Velocimetry, we measured a typical along-block downward velocity of $1.5 \, {\rm{cm/s}}$ at a distance of $10 \ \rm{mm}$ beneath the bottom interface of a caramel block dissolving in water and inclined of $60^{\circ}$ and of $2 \, {\rm{cm/s}}$ at a distance of $8 \ \rm{mm}$ beneath a salt block dissolving in water and inclined of $40^{\circ}$.  The differences in solute transport depending on the orientation of the block interface goes along with a different erosion velocity and shaping of the block walls. The block dissolves faster at interfaces facing downwards than at interfaces facing upwards. Moreover, while the walls facing upwards remain smooth, patterns form on the walls facing downwards. For the experiment shown in figure \ref{FigSeqCar}, longitudinal parallel stripes with an initial wavelength of 0.4 mm are observed a few second after immersion. These stripes extend on the whole length of the block after a given ``entry" distance from the top. Then, the wavelength increases with time, longitudinal stripes evolve into chevrons then into cup-like patterns named scallops by geologists, that propagate upwards, {\it{i.e.}} upstream.
After an hour, scallops are $7.6 \, \rm{mm}$ wide, $6.3 \, \rm{mm}$ long and propagate at a velocity of about $3 . 10^{-3} \, \rm{mm/s}$. They have the typical shape of rounded troughs and acute peaks with an asymmetry between the upstream and downstream sides.
Although patterns grow, evolve and propagate on the bottom interface, its overall inclination does not change over time (but in the vicinity of the top corner) and the global receding velocity is found remarkably constant through time as shown in figure \ref{FigHTCaramel}. A similar result is obtained in simulations of Rayleigh-B\'enard convection with a melting boundary, which show a constant receding rate of the mean position of the boundary while patterns are growing \cite{Favier_2019}.\\

Figure \ref{FigSeqSel} shows the evolution of patterns at the bottom interface of a salt block when immersed into water. The dissolution scenario we described applies for all the dissolution experiments we performed. However, the characteristic time and length scales depend on material, initial bath concentration or inclination of the block. We measure the initial wavelength of longitudinal stripes $\lambda_{\rm{s}}$, the characteristic growth time $T_{\rm{s}}$ for their formation and the global erosion velocity $\dot{h}$ of the block wall pointing downwards for blocks of caramel or salt for different inclinations and concentrations of solutions and for one block of plaster into water. Figure \ref{FigPar} shows early longitudinal stripes observed on the bottom wall of caramel blocks immersed in syrups of different concentration. The initial wavelength  as the characteristic time for their formation increase with the bath concentration. The interface recedes faster when the concentration of the bath solution is smaller. We also observe that the initial wavelength and growth time increase and the interface recedes slower when the inclination angle of the block increases.

\section{Scaling laws: a solutal Rayleigh-B\'enard instability \label{SecScal}}
The rate of dissolution (or precipitation) of a solid block depends on the difference between the concentration of the solution at the block interface and the saturation concentration. The simplest yet reasonable dependency assumes a linear relationship with a constant characteristic dissolution (or precipitation) velocity, which contains the chemical kinetics \cite{Lasaga_2014, Crank_1975}. The equation of conservation of the mass flux per unit of area reads:
\begin{equation}
\rho_{\rm{s}} \frac{\partial h}{\partial t} \vec{n}= \alpha \left( C_{\rm{i}} - C_{\rm{sat}} \right) \vec{n} 
\label{eqInter}
\end{equation}
where $\rho_{\rm{s}}$ is the density of the solid and $\alpha$ the characteristic dissolution velocity. $C_{\rm{i}}$ and $C_{\rm{sat}}$ are the mass concentrations of the solution at the interface and of the saturated solution, respectively. $\vec{n}$ is a unit vector, normal to the solid interface.

\begin{figure}[t]
 \begin{minipage}{0.48\linewidth}
  \includegraphics[width=\textwidth, clip]{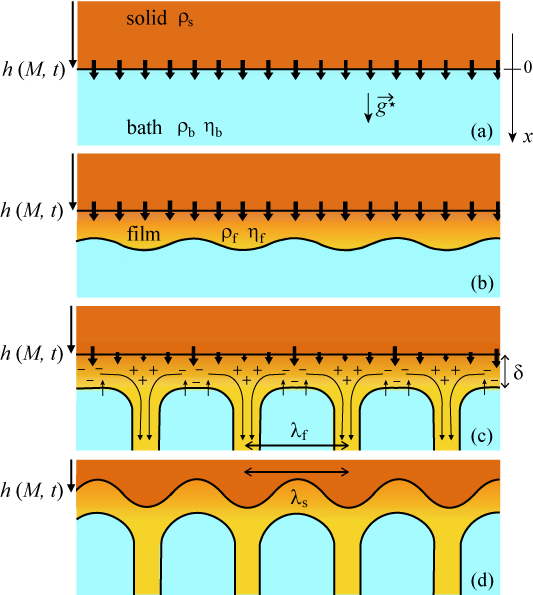}
\end{minipage}\hfill
  \begin{minipage}{0.48\linewidth}
  \caption{Sketch of the first steps of the dissolution process in a transverse plane. Time increases from top to bottom. The instability selects the thickness of the concentration layer and the wavelength of solute threads. It produces a modulation of the solute concentration at the interface, which shapes the dissolving block. The concentration at the interface of the block is larger above a thread $(+)$ than between threads $(-)$. The block dissolves faster where the concentration at the interface is smaller (eq. \ref{eqInter}). \label{FigShemaDissolution}} 
   \end{minipage}\hfill
     \end{figure}
For a flat and horizontal interface, if the dissolution process generates a density stratification that is buoyancy stable, only the diffusion drives the mass transport into the solution. 
Then, when putting a block in the solution, the concentration at the interface increases towards the saturation value on a characteristic time $D/\alpha^2$, where $D$ is the coefficient of diffusion of the solute. The receding velocity of the dissolving block decreases with time and for large times, the interface position withdraws as the square root of time \cite{Crank_1975}.

 If the dissolution process generates a density stratification that is buoyancy unstable, a completely different scenario occurs. When immersing the block, it starts dissolving and a front of dissolved medium moves away from the solid interface. On small lengths, close to the solid interface, the diffusion drives the transport. But beyond a characteristic length, the gravity takes over the transport. As a result, the concentration front destabilizes into plumes \cite{Philippi_2019} (see fig. \ref{FigShemaDissolution}). This instability is a solutal Rayleigh-B\'enard instability where concentration replaces temperature and the same scalings laws apply as previously observed by \cite{Schurr_1905, Tait_1989, Tait_1992a, Kerr_1994b, Sullivan_1996, Haudin_2014}.
Identically, it can be seen as a Rayleigh-Taylor instability of a thin film whose thickness and concentration are controlled by diffusion \cite{Kerr_1994b, Haudin_2014}.
In the classical Rayleigh-Taylor instability, the interface between two fluids of different density destabilizes when the heavier fluid is above the lighter one. The growth rate of perturbations is larger for small wavelengths. However, for non-miscible fluids, surface tension prevents the shorter wavelengths to grow and the most unstable wavelength is given by the capillary length. For a thin film when the surface tension goes to zero, the characteristic wavelength is given by the thickness of the film. H. R. Brown did the linear stability analysis in 2D for a horizontal thin film of a viscous liquid above an inviscid fluid and in contact with a solid wall at its upper limit \cite{Brown_1988}. The most unstable wavelength $\lambda_{\rm{f}}$ and its associated growth rate $\sigma_{\rm{f}}$ are \cite{Brown_1988, Limat_1993}: 
\begin{equation}
\lambda_{\rm{f}} \simeq 3 \, \delta \  \& \ \sigma_{\rm{f}} \simeq 0.32 \, \delta \Delta\rho \, g  / \eta_{\rm{f}},
\label{eqRT}
\end{equation}
where $\delta$ is the thickness of the destabilizing film, $\Delta \rho$ the apparent density, $\eta_{\rm{f}}$ the dynamic viscosity of the fluid in the film and $g$ the gravity acceleration. In Brown calculation there is no diffusion nor tangential stress at the interface because the outer fluid is inviscid. This last simplification holds for a viscous outer fluid when its dynamic viscosity is small compared to that of the fluid in the thin film and/or when the wavelength is large compared to the thickness of the film. Lister and Kerr extended the calculation for two fluids of different finite viscosities \cite{Lister_1989}. For two fluids with equaled dynamic viscosities, they found  $\lambda_{\rm{f}} \simeq 3.8 \, \delta$ \cite{Lister_1989}. Hereafter, we denote $R_{\lambda}$ the ratio between $\lambda_{\rm{f}}$ and $\delta$. Naturally, $\sigma_{\rm{f}}$ scales as the inverse of the characteristic settling time of a particle of diameter $\delta$ in terminal Stokes velocity over a length equal to its diameter.

The front between the layer of dissolved medium and the solvent is unstable when the gravity overcomes the diffusion to transport the dissolved mass and concentration. Thus, the thickness $\delta$ of the destabilizing film is such that the growth time scale of the instability $1/\sigma_{\rm{f}}$ equals the diffusion time over $\delta$: $\delta^2/ D$, which gives:
\begin{equation}
\delta \sim 1.5 \left( \frac{D \, \eta_{\rm{f}}}{\Delta \rho \, g}\right)^{1/3}.
\label{eqdeltaInsta}
\end{equation}
In a system where the outer bath is much deeper than $\delta$, the solute transport that the dissolution generates is always unstable and the onset of the instability corresponds to a constant value of the solutal Rayleigh number $Ra_{\rm{c}}$ \cite{Tait_1989,Sullivan_1996,Philippi_2019}:
\begin{equation}
Ra_{\rm{c}} = \frac{\Delta \rho \, g \delta^3}{D \eta_{\rm{f}}}.
\label{eqRac}
 \end{equation} 
Solute diffusion and viscosity stabilize the interface and tend to increase the film thickness $\delta$  while gravity drives the instability and tends to decrease $\delta$. The mean viscosity $\eta_{\rm{f}}$ and density $\rho_{\rm{f}}$ inside the film depend on the concentration of the solute-laden fluid.\\

We observe in experiments that the dissolution rate reaches a stationary state, {\it i.e.} the mass fluxes balance. In one dimension, the conservation of mass in a stationary regime reads:
\begin{eqnarray}
- \rho_{\rm{s}} \dot{h} & =& - D \, \frac{\partial C}{\partial x} \left( x \right) + C (x) \left[ v(x) - \dot{h} \right],
\end{eqnarray}
where  $x$ is the normal distance from the solid interface and $v$ the flow velocity in the laboratory frame (like $\dot{h}$). The diffusion coefficient $D$ is taken constant. In the solution, the flow and the diffusion of the solute concentration balance the dissolution of the solid, which is made of one species only. At the solid interface, the solute transport is diffusion-dominated while it is advection-dominated beyond the unstable solute layer, which gives three different equations for the dissolution flux:
\begin{eqnarray}
- \rho_{\rm{s}} \dot{h} & = & \alpha \left( C_{\rm{sat}} - C_{\rm{i}} \right) \label{FluxAlpha}\\
& \sim & D \frac{\left( C_{\rm{i}} - C_{\rm{b}} \right)}{\delta} - \beta \, \dot{h} \, C_{\rm{i}} \label{FluxD}\\
& \sim & \left( C_{\rm{f}} - \frac{\rho_{\rm{f}} - C_{\rm{f}}}{\rho_{\rm{b}} - C_{\rm{b}}} C_{\rm{b}} \right) \frac{4 (\rho_{\rm{f}}-\rho_{\rm{b}}) g \cos \theta \delta^4}{3 \eta_{\rm{f}} {\lambda_{\rm{f}}}^2}. \label{FluxG}
\end{eqnarray}

The first balance of solute flux per unit of surface is given by the kinetics of dissolution (eq. \ref{eqInter}).

The second line is the diffusion-dominated flux of mass at the solid interface. At the interface, the flow does not penetrate the solid but the normal velocity is not necessarily zero because the density of the solid and of the liquid may be different: $v(0)=\dot{h} (1-\beta)$, where $\beta$ is the ratio between the density of the solid and the density of the liquid at the interface ($\beta= \rho_{\rm{s}}/\rho_{\rm{i}}$) \cite{Quintard_1994, Wells_2011, Gagliardi_2018}.
The gravity takes over the transport a distance $\delta$ from the solid interface within flowing threads with a wavelength $\lambda_{\rm{f}}$ (see fig. \ref{FigShemaDissolution}). Thus, between threads and outside the mixture film, the concentration may be reduced to the one of the outer bath. The unstable solute layer assimilates to a boundary layer. Inside the boundary layer, the diffusion overcomes the advection so that the concentration decreases monotonically from the concentration $C_{\rm{i}}$  at the interface to the concentration $C_{\rm{b}}$ of the outer bath, almost linearly if $D$ does not depend on $C$ \cite{Philippi_2019}. The concentration gradient at the interface scales like $(C_{\rm{b}}-C_{\rm{i}})/\delta$.

The third line is the advection flux of mass at the boundary between the solute layer of thickness $\delta$ and the outer bath. It is buoyancy driven. The dissolution increases the concentration within the flowing film from the concentration of the outer bath $C_{\rm{b}}$ to the (mean) one of the film $C_{\rm{f}}$. The scaling of the flow velocity corresponds to a two dimensional lubrication flow of a fluid of density $\rho_{\rm{f}}$ and viscosity $\eta_{\rm{f}}$ inside a thin film of thickness $\delta$ confined between a rigid wall (the dissolving block) and an inviscid fluid of density $\rho_{\rm{b}}$ (the outer bath). The flow is driven by a characteristic pressure gradient $2 \delta (\rho_{\rm{b}} - \rho_{\rm{i}}) g^{\star} / \lambda_{\rm{f}}$ and fed half a thread (see fig. \ref{FigShemaDissolution}). Threads have a wavelength $\lambda_ {\rm{f}}$, which scales like $\delta$ (eq. \ref{eqRT}). $g^{\star}$ is the effective gravity for the growth of the instability. It should take into account the possible inclination of the dissolving interface. When the dissolving block makes an angle $\theta$ with the horizontal:  $g^{\star} = g \cos \theta$. It is the only way we take into account the inclination of the block, we neglect the longitudinal flow. Between threads, a flow of the bath solution at concentration $C_{\rm{b}}$ renews the water inside the solute layer. The factor in front of $C_{\rm{b}}$ comes from the conservation of water inside the solute layer. This ratio is always close to one. To obtain the scaling eq. \ref{FluxG}, we further assume that the thickness of threads is negligible compared to the distance $\lambda_{\rm{f}}$ between threads.\\

The equilibrium between the dissolution kinetics and the diffusive flux sets the concentration at the solid interface (eqs. \ref{FluxAlpha} and \ref{FluxD}). 
The concentration $C_{\rm{i}}$ has two asymptotic expressions depending on the ratio between the dissolution velocity $\alpha$ and the characteristic diffusion velocity over $\delta$: $D/\delta$. When the transport at the interface is limited by the diffusion, {\it{i.e.}} $D/(\alpha \delta) \ll 1$, the concentration at the interface is close to the saturation concentration. On the other hand, when the transport is limited by the dissolution kinetics, {\it{i.e.}} $D/(\alpha \delta) \gg 1$, the concentration at the interface is close to the concentration of the outer bath. For the dissolving materials we consider, the dissolution is limited by diffusion and we assume $C_{\rm{i}} = C_{\rm{sat}}$. The interested reader will find the scaling laws in the other limit in \cite{Philippi_2019}. 
We assume a linear concentration profile in the mixture film, so that the average concentration is $C_{\rm{f}} =(C_{\rm{sat}} + C_{\rm{b}})/2$. 
Finally, we assume a linear relationship between density and concentration, {\it{i.e.}} $\rho = \rho_0 +  C (\rho_{\rm{sat}} -\rho_0) / C_{\rm{sat}}$, where $\rho_0$ is the density of the solution when $C = 0$. Then, the density contrast reads $\rho_{\rm{f}}-\rho_{\rm{b}} = (\rho_{\rm{sat}}-\rho_{\rm{b}})/2$ and $\beta = \rho_{s}/\rho_{\rm{sat}}$. Combining equations \ref{FluxD} and \ref{FluxG} gives: 
\begin{equation}
\delta \sim \left( 3 {R_{\lambda}}^2 \right)^{1/3} \, \left[\frac{\rho_{\rm{s}}}{\rho_{\rm{s}} - \beta C_{\rm{sat}}} \ \frac{D \, \eta_{\rm{f}}}{(\rho_{\rm{sat}}-\rho_{\rm{b}}) \, g \cos \theta} \right]^{1/3}  \left(\frac{C_{\rm{sat}} - C_{\rm{b}}}{C_{\rm{sat}} - C_{\rm{b}} \frac{\rho_{\rm{sat}} - C_{\rm{sat}}}{\rho_{\rm{b}} - C_{\rm{b}}}}\right)^{1/3} \ \rm{and}
\label{eqdelta}
\end{equation}
\begin{equation}
\dot{h} \sim - \left( \frac{1}{3 {R_{\lambda}}^2} \right)^{1/3} \left[ \frac{D^2 \,  (\rho_{\rm{sat}}-\rho_{\rm{b}})^4 \, g \cos \theta}{\eta_{\rm{f}} \, \left( \rho_{\rm{s}} - \beta C_{\rm{sat}}\right)^2 \, \rho_{\rm{s}}}   \right]^{1/3}  \frac{C_{\rm{sat}}}{\rho_{\rm{sat}} - \rho_0} \left( \ \frac{C_{\rm{sat}} - C_{\rm{b}} \frac{\rho_{\rm{sat}} - C_{\rm{sat}}}{\rho_{\rm{b}} - C_{\rm{b}}}}{C_{\rm{sat}} - C_{\rm{b}}} \right)^{1/3},
\label{eqhPoint}
\end{equation}
where $R_{\lambda}$ is the ratio between the wavelength $\lambda_{\rm{f}}$ of the buoyant instability and the thickness $\delta$ of the solute layer as given by the analysis of the Rayleigh-Taylor instability (for non diffusing, homogeneous fluids) \cite{Brown_1988, Lister_1989}. 

The instability of the flow naturally reflects on the field of concentration at the solid interface. Above the flowing thread, the concentration should be maximum and the solid dissolution rate should be minimum (see fig. \ref{FigShemaDissolution}). The instability of the flow prints onto the solid and we expect that the first observed wavelength of patterns at the solid interface corresponds to the wavelength of the flowing threads: 
\begin{equation}
 \lambda_{\rm{s}} \sim \left( 3 {R_{\lambda}}^5 \right)^{1/3} \, \left[\frac{\rho_{\rm{s}}}{\rho_{\rm{s}} - \beta C_{\rm{sat}}} \ \frac{D \, \eta_{\rm{f}}}{(\rho_{\rm{sat}}-\rho_{\rm{b}}) \, g \cos \theta} \right]^{1/3} \left(\ \frac{C_{\rm{sat}} - C_{\rm{b}}}{C_{\rm{sat}} - C_{\rm{b}} \frac{\rho_{\rm{sat}} - C_{\rm{sat}}}{\rho_{\rm{b}} - C_{\rm{b}}}} \right)^{1/3}.
\label{eqLambdaS}
\end{equation}
Patterns grow because of a differential rate of dissolution imposed by a modulation of the solute concentration at the interface. The differential dissolution velocity should scale like the global erosion velocity $\dot{h}$ \cite{Philippi_2019} so that the characteristic growth time $T_{\rm{s}}$ for the formation of patterns, {\it{i.e.}} to observe a given or detectable aspect ratio, should scale like $\lambda_{\rm{s}} / \dot{h}$:  
\begin{equation}
T_{\rm{s}} \sim \left( 9 {R_{\lambda}}^7 \right)^{1/3}  \left[ \frac{{\eta_{\rm{f}}}^2 \left(\rho_{\rm{s}} - \beta C_{\rm{sat}} \right) {\rho_s}^2}{D \, \left(\rho_{\rm{sat}}-\rho_{\rm{b}} \right)^5 \, ( g \cos \theta)^2} \right]^{1/3}  \frac{\rho_{\rm{sat}} - \rho_0}{C_{\rm{sat}}} \left(\ \frac{C_{\rm{sat}} - C_{\rm{b}}}{C_{\rm{sat}} - C_{\rm{b}} \frac{\rho_{\rm{sat}} - C_{\rm{sat}}}{\rho_{\rm{b}} - C_{\rm{b}}}} \right)^{2/3}.
\label{Tpattern}
\end{equation}

These scaling laws are somewhat different than the previous scalings \cite{Sullivan_1996, Kerr_1994b, Philippi_2019}. Here, we derive the scaling laws from the conservation of mass assuming a stationary dissolution flux while previous scalings are based on a constant critical Rayleigh number (eq. \ref{eqRac}), {\it{i.e.}} derived assuming that the thickness $\delta$ of the boundary layer is such that the system is at the instability threshold. The two differences are the terms $(\rho_{\rm{s}}-\beta C_{\rm{sat}})$ and  $[C_{\rm{sat}} - C_{\rm{b}}(\rho_{\rm{sat}} - C_{\rm{sat}})/(\rho_{\rm{b}} - C_{\rm{b}})]/(C_{\rm{sat}} - C_{\rm{b}})  \equiv \Omega$ in ours scalings instead of $\rho_{\rm{s}}$ and $1$ respectively in previous scaling laws. The value of $\Omega$ increases with the bath concentration. It varies between $1$ and $1.13$ for salt and between $1$ and $2$ for caramel. This term does not change the quantitive values of the scaling laws very much. The term with the factor $\beta$ depends on the dissolving material. For usual natural materials, which have a very low saturation concentration ({\it{i.e.}} $C_{sat} \ll \rho_{\rm{sat}}$), the difference is incidental. It is not negligible for salt and caramel for which $\beta$, $C_{\rm{sat}}$, $\rho_{\rm{sat}}$ and $\rho_{\rm{s}}$ are $1.96$, $317 \, \rm{kg.m^3}$, $1197 \, \rm{kg.m^3}$ and $2348 \, \rm{kg.m^3}$ and $1.06$, $968 \, \rm{kg.m^3}$, $1450 \, \rm{kg.m^3}$ and $1540 \, \rm{kg.m^3}$, respectively.

\section{Results and discussion \label{SecRes}}
\begin{figure*}
\begin{center}
\includegraphics[width =\linewidth]{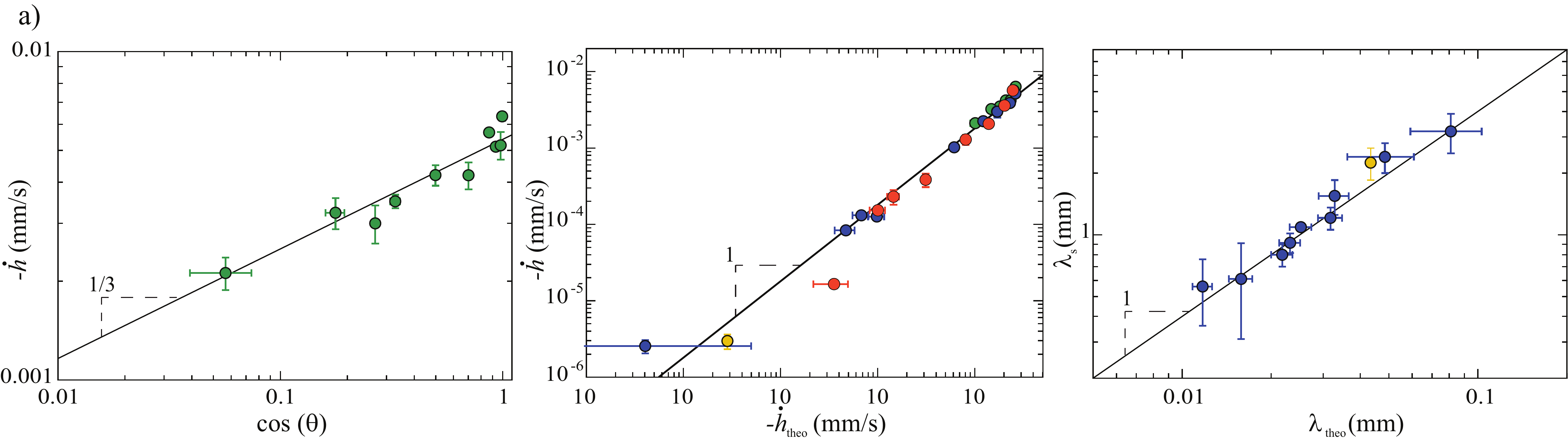}
\includegraphics[width =\linewidth]{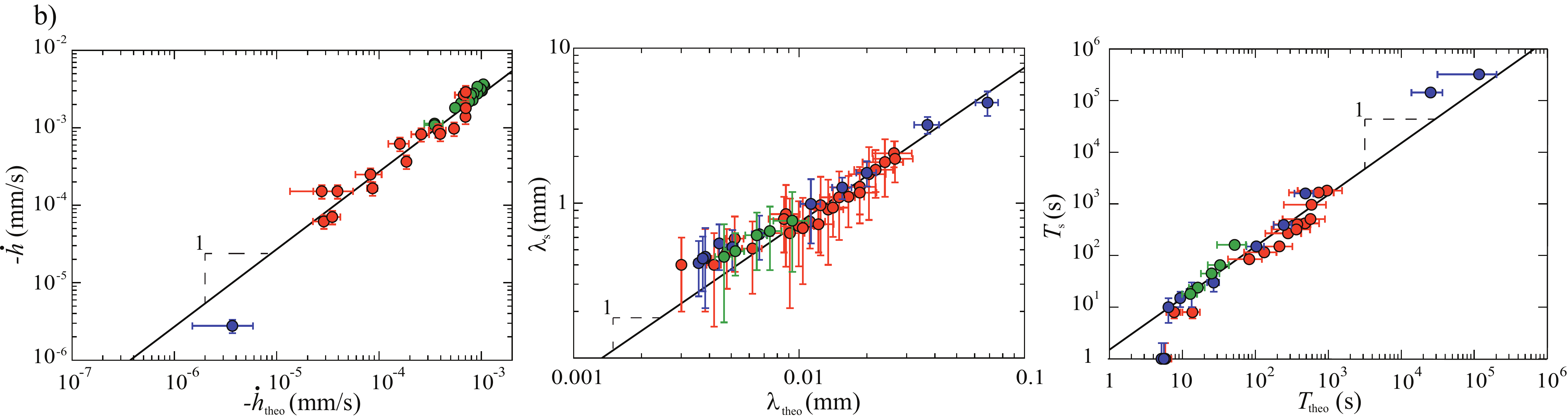}
\caption{Measurements of erosion velocity $\dot{h}$, first observed wavelength of patterns $\lambda_{\rm{s}}$ and growth characteristic time $T_{\rm{s}}$ as functions of scaling laws (without pre-factor) for experiments with salt and plaster blocks (a) and with caramel blocks (b). Vertical and horizontal error bars show the resolution or dispersion of measurements and the scaling confidence interval due to parameters uncertainties, respectively. Colors stand for the varied parameter. a) Green: pure water with $\theta$ ranging between $6$ and $87^{\circ}$,  orange: saline solutions of density $\rho_{\rm{b}}$ ranging between $1000$ and $1189 \, \rm{kg.m^{-3}}$ and $\theta \simeq 30^{\circ}$, blue: saline solutions of density $\rho_{\rm{b}}$ ranging between $997$ and $1197 \, \rm{kg.m^{-3}}$ and various inclinations $\theta$. The yellow point corresponds to gypsum into pure water with $\theta \simeq 52^{\circ}$. The line in the first graphic shows a power law with an exponent $1/3$. The lines in the two other graphics show relations of proportionality with constants equal to 0.18 for $\dot{h}$ and 40 for $\lambda_{\rm{s}}$. b - $\dot{h}$ graphic) Green: pure water with $\theta$ ranging between $0$ and $88^{\circ}$, orange: sucrose solution with density $\rho_{\rm{b}}$ ranging between $1000$ and $1320 \, \rm{kg.m^{-3}}$ and $\theta \simeq 73-75^{\circ}$, blue: caramel solution with density $\rho_{\rm{b}} = 1403 \, \rm{kg.m^{-3}}$ and $ \theta = 75^{\circ}$. 
b - $\lambda_{\rm{s}}$ $\&$ $T_{\rm{s}}$ graphics) Green: sucrose solution of density $\rho_{\rm{b}} = 1150 \, \rm{kg.m^{-3}}$ with $\theta$ ranging between $45$ and $85^{\circ}$, 
orange: sucrose solution with density $\rho_{\rm{b}}$ ranging between $1000$ and $1320 \, \rm{kg.m^{-3}}$ and $\theta \simeq 73 - 77^{\circ}$,
blue: caramel solution with density $\rho_{\rm{b}}$ ranging between $1000$ and $1430 \, \rm{kg.m^{-3}}$ and $\theta \simeq 70 - 75 ^{\circ}$. 
The coefficient of proportionality of black lines are 2.7 for $\dot{h}$, 75 for $\lambda_{\rm{s}}$ and 1.5 for $T_{\rm{s}}$.}
\label{FigDissScaling}
\end{center}
\end{figure*}

To calculate these scaling laws one has to chose values for the coefficient of diffusion $D$ and for the viscosity $\eta_{\rm{f}}$. We take the value of $\eta_{\rm{f}}$ that corresponds to the assumed mean concentration in the film, {\it{i.e.}} for $(C_{\rm{sat}}+C_{\rm{b}})/2$. We measured the viscosity as a function of the solute concentration for solutions of caramel, saccharose and salt. Because the saturation concentration of plaster in water is very weak, we take the viscosity of water for the plaster experiment. For $D$, we take the value that corresponds to the saturated concentration, {\it{i.e.}} the expected one at the block interface. It is determined from handbooks for salt and plaster and from the tabulated value at zero concentration for saccharose, which we extrapolate to the value at saturation using the Stokes-Einstein relation. These choices are not critical in the case of plaster, whose saturation concentration is very weak and in the case of salt, for which $D$ barely depends on concentration and the viscosity changes by a factor of two between the pure water and the saturated brine \cite{Vitagliano_1956}. However, these choices are not insignificant in the case of caramel ; the viscosity value of the saturated syrup is $4 \times 10^{4}$ times larger than the value of water. Most previous studies do not address the values of these coefficients, which is justified for materials with a small value of $C_{\rm{sat}}$ without impact on these parameters.

Figure \ref{FigDissScaling} compares the experimental results to the scaling laws without pre-factors, which include $R_\lambda$. The various dependencies of the erosion velocity and the initial wavelength of patterns with the density contrast, the block inclination and, for caramel, the viscosity through the solute concentration, are verified over several decades. We find a pre-factor of about 0.18 for the dissolving velocity of salt blocks, which is very close to the expected order of magnitude ($\sim 0.29$ with $R_{\lambda} = 3.8$, eq. \ref{eqhPoint}).
For the dissolving velocity of caramel blocks the pre-factor is larger and equals $2.7$ when the expected one is about 0.34 (with $R_{\lambda} = 3$). The choices we do for the values of $D$ and $\eta_{\rm{f}}$ 
may explain the different pre-factors for experiments with salt and caramel. 
Note that the scaling laws compare well to the measurements even if derived neglecting the solute flow along the dissolving block and the dynamics of patterns at the interface. Patterns are ignored and the inclination of the block is taken into account only through the projection of the gravity, which gives good agreement with experimental data (fig. \ref{FigDissScaling}). This is a noteworthy result for predictions in natural environments.

A horizontal interface dissolves faster than an inclined one. This result suggests that a polygonal or rounded object should not keep a constant shape while dissolving. This is different from recent observations of Davies Wykes {\it et al.}, who report that the gravity-driven dissolving velocity of a block does not depend on its inclination \cite{Wykes_2018}. This may be ascribed to the limited range of their experiments and the weak dependency at small angles ($-\dot{h} \propto \cos ^{1/3} \theta$, eq. \ref{eqhPoint}) and/or to the relative small size of objects they consider.  The instability fully develops after an entry length (see fig. \ref{FigSeqCar}) 
and the differences in macroscopic receding velocity should affect the overall shape of the object if the latter is way larger than the size of patterns.

Our experiments where blocks are inclined clearly show a characteristic wavelength for the initial stripes that longitudinally extend all along the block. The initial wavelength of patterns we measure agree with the scaling law of the most unstable wavelength of the flow instability (fig. \ref{FigDissScaling}, eq. \ref{eqLambdaS}). Like for the erosion velocity, the observed pre-factor is closer for experiments with salt ($\sim 40$ instead of $\sim 13$) than for experiments with caramel ($\sim 75$ instead of $\sim 8.8$). Note that measurements may deviate from the scaling law at small values. However, this could be ascribed to experimental issues like an initial roughness of the solid surface or to the difficulty of measuring the very first wavelength when the patterns evolve rapidly. 

Figure \ref{FigDissScaling} shows that the characteristic growth time of patterns on caramel blocks properly scales like $\lambda_{\rm{s}} / \dot{h}$ (eq. \ref{Tpattern}). This raises the question of the pattern growth rate. Most of instabilities grow exponentially at first in the linear regime. An exponential growth demands a coupling between the cause and the effect. Such a coupling is not obvious here where the patterning of the block comes from the buoyant instability of the dissolution flow. However,  the flow instability can imprint onto the dissolving solid only if the positions of flowing threads are stabilized. A possible interaction between the patterns and the flow could rely on the spatial stabilization of dissolution plumes. The position of plumes may be all the more stable (stationary) that a dissolution peak is developed overhead. It is worth noting that the wavelength of initial patterns does not seem as well-defined as in our experiments with inclined blocks when interfaces are horizontal \cite{Sullivan_1996}. In our experiments, the dissolution flow along the block could stabilize the transverse position of early plumes and limit their interactions. Indeed, the longitudinal extension of early stripes is striking (fig. \ref{FigPar}). Note that an outer flow, transverse to the growth direction of the instability, reduces the flow pattern to two dimensions (longitudinal rolls) in Rayleigh-B\'enard experiments with small gaps \cite{Kelly_1994, Pabiou_2005}. In our experiments the transverse-to-growth-direction flow is the dissolution flow along the inclined block itself. It is the opposite case of the Rayleigh-B\'enard vortices in granular flows where it is the flow that ``heats" the granular medium \cite{Forterre_2001}.   
Anyway, without coupling between the amplitude of stripes and their growth velocity, the amplitude of early stripes would grow linearly in time. Numerical simulations could address this question effectively.\\

\begin{figure*}[t]
\begin{center}
\includegraphics[width =\linewidth]{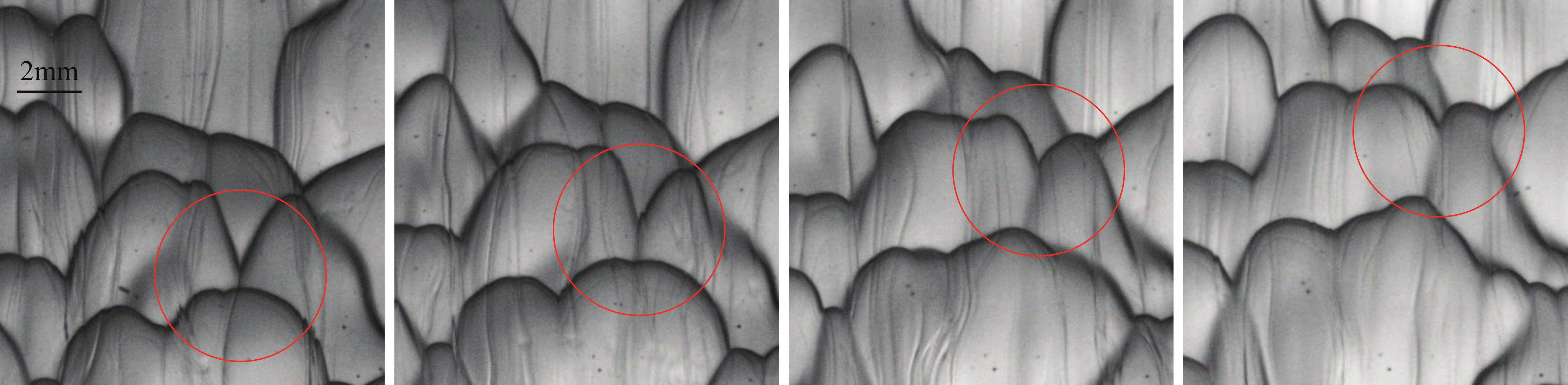}
\caption{Pattern dynamics. Bottom view of a caramel block dissolving into water.  A V-shape pointing downstream vanishes. The block makes an angle of $60^{\circ}$ with the horizontal. Gravity points towards the bottom of images. There is a 6 minutes time delay between snapshots. First image is taken 80 minutes after immersion.}
\label{FigV}
\end{center}
\end{figure*}

We observe experimentally that the initial longitudinal stripes do not last and turn into scallop-like patterns that deepen, enlarge, elongate, and propagate upstream.
This is definitely a true interaction between the patterns and the dissolution flow.
Figure \ref{FigCarCote} shows that solute plumes preferentially detach the boundary layer from peaks. This explains the upstream propagation of patterns, the reorientation of stripes and the final concave shape of scallops with acute peaks and rounded troughs as shown in figure \ref{FigSeqSel}. The buoyancy driven dissolution tends to form and preserves singularities. Following the solute flow along the solid interface, the concentration is larger, closer to the saturation value, and the concentration gradient is smaller because of a thicker boundary layer, upstream from the peak, before the plume detaches, than downstream. Thus, nearby a peak where a plume detaches, the dissolution flux is larger downstream than upstream from the peak. This differential dissolution makes that the peak (and patterns) propagate upstream. Crests channelizing the solute flow, the initial longitudinal stripes are unstable because of an inhomogeneous propagation velocity, transverse or longitudinal. The propagation velocity is maximum where plumes detach because the spatial variation of the solute flux is maximum there. If we consider for example that a stripe is not perfectly longitudinal but makes a small angle with the downstream flow, it will break and reorientate where plumes detach.  The image sequence shown in fig. \ref{FigV} illustrates the propensity of patterns to orientate transversally. 

\section{Analogy with the formation of Penitentes and sublimation patterns \label{SecPenitentes}}

Penitentes are ice or snow spikes that form on high altitudes glaciers in sublimation conditions, and have recently been suspected on other planetary bodies \cite{Moores_2017, Hobley_2018}. They point in the direction of the sun, which suggests that the sublimation is driven by the heat flux resulting from the light absorption by the ice. Previous studies proposed a coupling between the ice topography and the sublimation rate to explain the pattern formation \cite{Lliboitry_1954, Betterton_2001, Bartels_2012, Cathles_2014, Claudin_2015}. Because of a curvature effect, a trough receives more light than a peak and a perturbation of the topography grows. The corresponding linear stability analysis, where heat and mass diffuse but without advection, does not predict the selection of an initial wavelength \cite{Claudin_2015}. Claudin {\it{et al.}} propose that the wavelength may be selected by the boundary layer of an external air flow above the ice surface \cite{Claudin_2015}. However, Bergeron {\it{et al.}} report that an air flow above the snow surface prevents the formation of penitentes in their experiment \cite{Bergeron_2006}.
 
We propose that penitentes firstly form by differential sublimation because of a buoyancy instability in the flow of moist air above the ice surface, like for the dissolution patterns we report. The initial ice peaks would then not form because of the coupling between the topography and the radiative heat flux but would simply be a print of the air flow instability. The buoyancy instability would be driven by moisture because moist air is lighter than dry air. 

The ice sublimates because the partial pressure of water vapor in the air is lower than the equilibrium vapor pressure.
At the ice interface, the air is almost saturated in water vapor with a concentration $C_{\rm{sat}}$, which depends on the temperature.
The incident light brings energy to sublimate the ice but also to heat the (moist) air at the ice interface. The warming of air at the ice interface enhances the instability because the equilibrium vapor pressure of water in the air increases with temperature and so does the buoyancy. Thereby, the energy balance must be resolved concurrently with the equations of mass conservation.
The incident light brings a characteristic heat flux, which is absorbed by the ice. In a steady state of sublimation, the absorbed heat at the interface diffuses in the ice, diffuses and convects in the air and provides the energy for the sublimation.
The balance between the sublimation fluxes derived from the conservation of mass and derived from the conservation of energy would allow the sublimation rate, the thickness of the buoyancy unstable layer of moist air (and the wavelength between threads) together with the temperature of the air at the ice / air interface to be calculated.

A full analysis of this problem is beyond the scope of our article. However, we can estimate the first wavelength of buoyancy driven penitentes for given temperatures of the moist air layer and of the dry ambient air, which set the parameters value as the saturation concentration of the water vapor. For a dry atmosphere at $-20 ^{\circ}\rm{C}$, if the ice interface is heated up to $-5^{\circ}\rm{C}$, we find $\lambda_{\rm{s}} \simeq 32 \, \rm{mm}$ according to eq. \ref{eqLambdaS} with the pre-factor we measure for salt dissolution (see fig. \ref{FigDissScaling}) \footnote{This value corresponds to $D = 2.14 \times 10^{-5} \rm{m^{2}.s^{-1}}$, $\eta_{\rm{f}} = 1.65 \times 10^{-5} \rm{Pa.s}$, $\rho_{\rm{sat}} = 1.314 \rm{kg.m^{-3}}$, $\rho_{\rm{b}} =1.394 \rm{kg.m^{-3}}$.}. It compares to the typical centimeter scale observed in experiments \cite{Bergeron_2006}. 
Although it might be annihilated in windy conditions, the buoyancy driven sublimation and evaporation might make non negligible contributions to water budgets \cite{Shahidzadeh_2006, Dehaeck_2014, Carrier_2016, Dollet_2017}.  

\section{Conclusion, free convection and patterns in natural environment
\label{SecConclu}}
Our experimental observation of dissolution of various solids in aqueous solutions evidences the succession of two pattern formation mechanisms.
First, a longitudinal instability develops with the appearance of stripes carved on the surface and spaced out by a well-defined wavelength. We source it to the buoyancy hydrodynamical instability due to the density stratification that the dissolution generates. It ends up to the emission of solute plumes spatially separated. We show that the inclination of the dissolving blocks stabilizes the location of the plumes in order to print coherently the block surface. We confirm that this observed primary pattern follows scaling laws based on the hydrodynamics only.
At later times the stripes eventually turn into scallops. This second mechanism involves strong interactions between the flow and the topography. The final shape displays very stereotyped scallops, similar to some observed in nature, advocating for the importance of the role played by buoyancy in natural systems.\\

Underground cavities shaped by free convection in nature are reported in several environments including salt, gypsum and limestone rocks. They occur as dissolutional chambers or vertical breakdown structures and are confined systems \cite{Klimchouk_2007}. The dissolutional chambers devellop upwards and the buoyancy-driven flow shapes the cavity into an upsidedown triangular prism or cone with a flat horizontal ceiling \cite{Gechter_2008}. This particular shape is another indication that free convection increases the dissolution rate. Onto the top walls, centimeter cups with sharp edges named dissolution pits are observed \cite{Kempe_1972, Kempe_1996}. 
In our experiments with inclined blocks, the morphology of the dissolution pits is reminiscent of the scallop like patterns observed in unconfined systems such as caves or icebergs. However, these lasts are likely produced by the turbulent flow of a ground river or an ocean current. An externally imposed flow should prevail on the buoyancy to drive the dissolution when the turbulence of the flow decreases the thickness of the concentration boundary layer below the one imposed by the solutal Rayleigh B\'enard instability (eqs. \ref{eqdeltaInsta}, \ref{eqdelta}). 
Regarding heat transfer, numbers of studies address these questions \cite{Bergman_2011, Pirozzoli_2017}.

Scallops shaped by an external flow have been successfully reproduced in laboratory flumes with a turbulent flow over beds of plaster \cite{Allen_1971a, Blumberg_1974, Villien_2005} or ice \cite{Ashton_1972, Gilpin_1980, Bushuk_2019}. These studies report that mature patterns in experiments have constant amplitude and longitudinal wavelength (periodicity in the stream direction), which scales as one over the flow velocity. In the field, scallops could thus be used as proxies of the flow \cite{Curl_1966, Curl_1974}.
Theoretically, Hanratty and Claudin {\it{et al.}} studied the pattern formation on a bed sheared by a turbulent flow in the limit of deep flows  \cite{Hanratty_1981, Claudin_2017}. Patterns arise from the coupling between the dissolving topography and the turbulent viscosity at the solid interface, which increases the solute transport \cite{Claudin_2017}.
Interestingly, the positive feedback is restricted to small values of the hydrodynamic roughness of the flow, and to a narrow range of wavelengths, which scale as one over the characteristic velocity of the flow \cite{Hanratty_1981, Claudin_2017}. Considering the relative small aspect ratios of scallops observed in experiments and in the field, this result obtained in the linear regime could explain the longitudinal wavelengths and amplitudes of mature scallops. The 3-D shape of these scallops remains an open question. 
In our experiments,  the convective solute flow driven by the dissolution of inclined blocks shapes bed forms through differential dissolution. Initial bedforms are longitudinal ridges, which turn into 3-D scallop-like patterns. Unlike the scallops in turbulent flows, the amplitude, width and length of the free dissolution scallops are not constant nor saturate in our experiments, but increase with time. Moreover, the free dissolution scallops propagate upstream while scallops in turbulent flow propagate downstream \cite{Curl_1966, Blumberg_1974, Allen_1971a,Villien_2005, Bushuk_2019}. This difference in propagation direction reflects a different shift between the topography and the erosion rate.
The erosion must be maximum on the lee side of free dissolution scallops, and on the stoss side of scallops in a turbulent flow. If these last ones propagate downstream but do not grow in size, the erosion is maximum at mid slope. Despite the different space shifts between the erosion rate and the topography, these two patterns look similar. They exhibit sharp peaks and asymmetric profiles with a larger slope on the lee side than on the stoss side. The peaks may however be more acute in our experiments with a  rather concave shape upstream while the stoss side of the peaks appear more rounded in flows experiments \cite{Curl_1966, Blumberg_1974, Allen_1971a}.
The overall resemblance between in flow and free dissolution scallops incites one to link some aspects of their morphogenesis. In both cases, the transverse topography could be due to an instability in the dissolution flow only, while a true feedback between the topography and the solute flow should be responsible for the longitudinal topography. The first observed patterns could be longitudinal ridges in both cases and they could trigger the transverse instability.

\begin{acknowledgments}
We thank Cyril Ozouf who did some experiments during an internship in our laboratory and Marc Durand for fruitful discussions. We thank J\'er\^ome Jovet for his help to do Schlieren imaging and Imane Boucenna for sharing her rheometer. This research was funded by the french Agence Nationale de la Recherche (grants ANR-12-BS05-0001/EXO-DUNES and ANR-16-CE30-0005/ERODISS).
\end{acknowledgments}

\bibliography{DissolutionArXiv}

\end{document}